\documentclass[a4paper]{article}

\usepackage{authblk}
\usepackage[utf8]{inputenc}
\usepackage[T1]{fontenc}
\usepackage{pslatex}
\usepackage[english]{babel}
\usepackage{fancyhdr}
\usepackage{hyperref}
\usepackage{amsmath,amssymb,amsfonts}
\usepackage{graphicx}
\usepackage{amsmath}
\usepackage{amsthm}
\usepackage{amsfonts}
\usepackage{amssymb}
\usepackage{algorithm}
\usepackage{microtype}
\usepackage{rotating}
\usepackage{todonotes}
\usepackage{tikz}
\usepackage{xparse}
\usepackage{appendix}
\usepackage{url}
\usepackage{tabularx}
\usepackage{paralist}
\usepackage{booktabs}
\usepackage{multirow}
\usepackage{subcaption}
\usepackage{cancel}
\usetikzlibrary{calc,shapes.callouts,shapes.arrows}
\usetikzlibrary{automata}
\usetikzlibrary{positioning}
\usetikzlibrary{decorations.pathreplacing}
\usetikzlibrary{decorations.pathmorphing}
\usetikzlibrary{arrows}
\usetikzlibrary{shapes}

\newcommand{\N}{\mathbb{N}}

\newcommand{\F}{\mathbb{F}}

\newtheorem{definition}{Definition}
\newtheorem{lemma}{Lemma}
\newtheorem*{problem*}{Problem}
\newtheorem{remark}{Remark}

\providecommand{\keywords}[1]{\textbf{\textit{Keywords }} #1}

\begin{document}

\title{Exhaustive Generation of Linear Orthogonal Cellular Automata\thanks{This paper is a shortened version of~\cite{formenti22}, which is currently under submission at \emph{Cryptography and Communications.}}}

\author[1]{Enrico Formenti}
\author[2]{Luca Mariot}
	
\affil[1]{{\normalsize Laboratoire d’Informatique, Signaux et Syst\`{e}mes de Sophia-Antipolis (I3S), Universit\'{e} C\^{o}te d’Azur, 2000, route des Lucioles - Les Algorithmes, b\^{a}t. Euclide B, 06900 Sophia Antipolis, France} \\
	
	{\small \texttt{enrico.formenti@unice.fr}}}

\affil[2]{{\normalsize Semantics, Cybersecurity and Services Group, University of Twente, Drienerlolaan 5, 7522NB, Enschede, The Netherlands} \\
	
	{\small \texttt{l.mariot@utwente.nl}}}
	
\maketitle

\begin{abstract}
We consider the problem of exhaustively visiting all pairs of linear cellular automata which give rise to orthogonal Latin squares, i.e., linear Orthogonal Cellular Automata (OCA). The problem is equivalent to enumerating all pairs of coprime polynomials over a finite field having the same degree and a nonzero constant term. While previous research showed how to count all such pairs for a given degree and order of the finite field, no practical enumeration algorithms have been proposed so far. Here, we start closing this gap by addressing the case of polynomials defined over the field $\F_2$, which corresponds to binary CA. In particular, we exploit Benjamin and Bennett's bijection between coprime and non-coprime pairs of polynomials, which enables us to organize our study along three subproblems, namely the enumeration and count of: (1) sequences of constant terms, (2) sequences of degrees, and (3) sequences of intermediate terms. In the course of this investigation, we unveil interesting connections with algebraic language theory and combinatorics, obtaining an enumeration algorithm and an alternative derivation of the counting formula for this problem.
\end{abstract}


\section{Introduction}
\label{sec:intro}
Orthogonal Cellular Automata (OCA) are pairs of CA whose global rules form orthogonal Latin squares, introduced by Mariot et al. at AUTOMATA 2016~\cite{mariot16}. As such, OCA have several practical applications in cryptography and coding theory, including threshold secret sharing schemes~\cite{mariot17,mariot18}, primary constructions for partial spread bent functions~\cite{gadouleau20,gadouleau23}, polynomial codes~\cite{mariot22a} and pseudorandom number generators with guaranteed diffusion properties~\cite{mariot21,mariot22}. There is an algebraic characterization of OCA when the underlying local rules are linear, which relates the orthogonality of the resulting Latin squares to the coprimality of the polynomials induced by the rules~\cite{mariot20}. Thus, the question of  enumerating all pairs of linear OCA boils down to enumerating a specific kind of coprime polynomials over a finite field---namely, those pairs where both polynomials have the same degree and a nonzero constant term. 

In general, coprime polynomials over finite fields have been studied in depth in the literature, also due to their numerous cryptographic and coding-theoretic applications~\cite{shparlinski13,coppersmith84,fitzpatrick95}. Most of the related works in this area concern counting results with no restrictions on the constant terms nor the degree of the involved polynomials~\cite{corteel98,reifegerste00,benjamin07}. Benjamin and Bennett~\cite{benjamin07} described a simple bijection between coprime and non-coprime pairs for polynomials over the field $\F_2$. Such a bijection is defined by using Euclid's algorithm and its application in reverse order, named dilcuE's algorithm in~\cite{benjamin07}. On the other hand, there are fewer works explicitly considering the exhaustive enumeration of coprime polynomial pairs~\cite{fragneto05}.

In this paper, we show how to exhaustively generate all linear binary OCA by enumerating all coprime polynomial pairs over the finite field $\F_2$, where both polynomials have the same degree and a nonzero contant term. To this end, we take inspiration from Benjamin and Bennett's bijection and show that the sequences of quotients visited by dilcuE's algorithm for these coprime pairs are characterized by three independent components: the \emph{constant terms}, the \emph{degrees} and the \emph{intermediate terms}. From these results, we present the pseudocode of a combinatorial algorithm that generates all pairs of coprime polynomials of degree $n$ and nonzero constant term. Further, we also provide an alternative derivation of the recurrence proved in~\cite{mariot20} to count all such pairs for $q=2$.

\section{Background Definitions}
\label{sec:bg}
Cellular Automata (CA) are usually regarded as a parallel computational model or a particular kind of discrete dynamical systems. Here, we interpret CA as algebraic systems instead, specifically as a type of vectorial mappings composed of uniform local coordinate functions. Since we are interested only in binary CA, we can define them in terms of vectorial Boolean functions:
\begin{definition}
	\label{def:ca}
	A Cellular Automaton (CA) of length $m \in \N$, diameter $d \le m$, and local rule $f: \F_2^d \to \F_2$ is a map $F: \F_2^m \to \F_2^{m-d+1}$ defined for all $x \in \F_2^m$ as:
	\begin{equation}
		\label{eq:ca}
		F(x_0,\ldots,x_{m-1}) = (f(x_0,\ldots, x_{d-1}),f(x_1,\ldots, x_{d}),\ldots,f(x_{m-d},\ldots,x_{m-1})) \enspace .
	\end{equation}
\end{definition}
Hence, each output coordinate $i \in \{0,\cdots, m-d\}$ of a CA $F$ is defined as the application of the local rule $f$ over the neighborhood composed by the $i$-th input cell and the $d-1$ cells to its right.

Next, we give the formal definition of orthogonal Latin squares:
\begin{definition}
	\label{def:ols}
	Let $[N] = \{1,\cdots, N-1\}$ for all $N \in \N$. A Latin square of order $N$ is a $N \times N$ matrix $L$ with entries over $[N]$ such that each row and each column of $L$ is a permutation of $[N]$. Two Latin squares $L_1,L_2$ of order $N$ are called orthogonal if the function $H: [N]\times [N] \to [N] \times [N]$ defined for all $(i,j) \in [N] \times [N]$ as $H(i,j) = (L_1(i,j), L_2(i,j))$ is bijective.
\end{definition}
Intuitively, two Latin squares are orthogonal if and only if their \emph{superposition} yields all pairs in the Cartesian product $[N] \times [N]$ exactly once.

Suppose now that $m = 2(d-1)$ and $F: \F_2^{2(d-1)} \to \F_2^{d-1}$ is a CA equipped with a \emph{bipermutive} local rule, i.e. $f: \F_2^d \to \F_2$ is of the form $f(x_0,\cdots, x_{d-1}) = x_1 \oplus g(x_1,\cdots,x_{d-2}) \oplus x_{d-1}$, where $g: \F_2^{d-2} \to \F_2$ is any Boolean function over the $d-2$ central cells. In this case, Mariot et al.~\cite{mariot16} showed that $F$ defines a Latin square $L_F$ of order $N = 2^{d-1}$. The idea is to encode blocks of $d-1$ bits in their decimal form. Then, the left and right $(d-1)$-bit blocks composing the input vector of $F$ are used to index respectively the row and the column coordinates of $L_F$, while the output $(d-1)$-bit block of $F$ represents the entry to be placed at those coordinates.

Let us further assume that the local rule is \emph{linear}. This means that $f: \F_2^d \to \F_2$ is defined as $f(x_0,\cdots,x_{d-1}) = x_0 \oplus a_1x_1 \oplus \cdots \oplus a_{d-2}x_{d-2} \oplus x_{d-1}$, where $a_i \in \F_2$ for all $i \in \{1,\cdots, d-2\}$. In other words, $f$ is an $\F_2$-linear combination of the neighborhood cells, with the property that the leftmost and righmost cells are always XORed. In this case, a polynomial of degree $n = d-1$ in the ring $\F_2[X]$ can be naturally associated to the local rule:
\begin{equation}
	f \mapsto P_f(X) = 1 + a_2X + \cdots + a_{n-1}X^{n-1} + X^n \enspace . 
\end{equation}
Hence, we use the coefficients of the linear local rule to index the increasing powers of the indeterminate $X$ in the polynomial $P_f$. The polynomial is thus monic of degree $n$ and with a nonzero constant term.

In~\cite{mariot16}, the authors showed that two CA respectively defined by linear bipermutive local rules $f,g:\F_2^d \to \F_2$ give rise to a pair of orthogonal Latin squares if and only if their associated polynomials are relatively prime, i.e. if and only if $\gcd(P_f,P_g) = 1$. Therefore, the problem of counting and enumerating all pairs of linear Orthogonal Cellular Automata (OCA) of diameter $d$ is equivalent to counting and enumerating all pairs of coprime polynomials of degree $n=d-1$ with a nonzero constant term. The counting question has already been settled in~\cite{mariot20} for any finite field order $q$, where $q$ is a power of a prime. 

\section{Problem Statement}
\label{sec:prob}
In this paper, we focus on the enumeration problem for the binary case $q=2$. To this end, we define the set $S_n$ for all $n \in \N$ as:
\begin{equation}
	\label{eq:sn}
	S_n = \{f \in \F_2[x]: x^n + a_{n-1}x^{n-1} + \ldots + a_1x + a_0: a_0 = 1 \} \enspace ,
\end{equation}
that is, $S_n$ is the set of binary polynomials of degree $n$ with nonzero constant term. Further, let $A_n$ and $B_n$ be respectively defined for $n \in \N$ as:
\begin{equation}
	A_n = \{(f,g) \in S_n^2: \gcd(f,g) = 1\} \enspace , \enspace B_n = \{(f,g) \in S_n^2: \gcd(f,g) \neq 1\} \enspace .
\end{equation}
Thus, $A_n$ and $B_n$ are the sets of pairs of polynomials of degree $n$ and nonzero constant terms that are respectively coprime and non-coprime. Clearly, it holds $A_n \cup B_n = S_n^2$ and $A_n \cap B_n = \varnothing$. We now formally state the problem addressed in this paper. Given $n \in \N$, we aim to:
\begin{compactitem}
	\item[(i)] \emph{Enumeration}: Find an algorithm to exhaustively generate all elements of $A_n$.
	\item[(ii)] \emph{Counting}: Derive a formula for $a_n=|A_n|$.
\end{compactitem}
This problem has already been solved in~\cite{benjamin07} for the case of generic polynomials of degree $n$ (i.e. with unconstrained constant terms). The idea is to define a bijection between the sets of coprime and non-coprime pairs as follows:
\begin{compactenum}
	\item Apply Euclid's algorithm to a non-coprime pair $(f,g)$.
	\item Change the last remainder from $0$ to $1$, and invert Euclid's algorithm (i.e. apply dilcuE's algorithm) using the same sequence of quotients computed for $(f,g)$ in reverse order. 
	\item By construction, the pair $(f',g')$ at the end of dilcuE's algorithm is coprime.
\end{compactenum}
The crucial remark in the above procedure is that \emph{the family of all sequences of quotients defines a bijection between coprime and non-coprime pairs}. The last remainder, either $1$ or $0$, defines whether the pair is respectively coprime or not.

Notice that if one employs this procedure starting from a non-coprime pair $(f,g) \in B_n$, in general the resulting coprime pair $(f',g')$ will not belong to $A_n$, i.e. either $f'$ or $g'$ could have a null constant term. Nonetheless, since $f'$ and $g'$ are coprime, it cannot be the case that both of them have a null constant term, otherwise they would have a factor $x$ in common. Therefore, we need to see how changing the last remainder in Euclid's algorithm affects the constant terms of the intermediate remainders, and thus the constant terms of $f'$ and $g'$.

Let us start with the following remarks:

\begin{remark}
	\label{rem:first-last}
	Let $(f,g)$ be two polynomials of degree $n$. Then:
	\begin{compactitem}
		\item[(i)] The first quotient obtained from Euclid's algorithm is always $q_1 = 1$. Indeed, since $f$ and $g$ have both degree $n$, the long division stops immediately after dividing $x^n$ by $x^n$.
		\item[(ii)] Suppose that $\gcd(f,g) = 1$. Then, when the last pair of remainders is $(r_k(x), 1)$, if we apply Euclid's algorithm for one further step we will always obtain the pair $(1,0)$ with quotient $r_k(x)$. In fact we can write the division of $r_k(x)$ and $1$ as $r_k(x) = r_k(x) \cdot 1 + 0$.
	\end{compactitem}
\end{remark}

Suppose that $q_1, q_2,\cdots, q_k$ is a sequence of quotients that yield a coprime pair $(f,g) \in A_n$ when applied in reverse order from $(1,0)$ through dilcuE's algorithm. We denote these quotients as:
\begin{align*}
	q_1 &\to \overbrace{x^{d_1}}^{\textrm{degrees}} \, + \overbrace{q_{1,d_1-1}x^{d_1-1} + \cdots + q_{1,1}x}^{\textrm{intermediate terms}} + \overbrace{s_1}^{\textrm{constant terms}} \\
	q_2 &\to \ \ x^{d_2} \ \: + q_{2,d_2-1}x^{d_2-1} + \cdots + q_{2,1}x + \ \ \ \ \ \ s_2 \\
	\vdots &\to \ \ \ \vdots \ \ \ \, \, \, + \ \ \ \ \vdots \ \ \ \ \ \ \ \ \ \ \, \; + \cdots + \ \ \vdots \ \ \ \, + \ \ \ \ \ \ \ \vdots \\
	q_k &\to \ \ x^{d_k} \ \: + q_{k,d_k-1}x^{d_k-1} + \cdots + q_{k,1}x + \ \ \ \ \ \ s_k \\
\end{align*}
with $d_1,\cdots,d_k \in \N$ being the \emph{degrees} of the quotients, $q_{i,j} \in \F_2$ the coefficients of the \emph{intermediate terms}, and $s_i \in \F_2$ the \emph{constant terms}.

Notice that a sequence of quotients is defined by independently choosing each of these three elements. Since our goal is to obtain a pair $(f,g) \in A_n$, the following two constraints hold:
\begin{compactitem}
	\item The sum of the degrees $d_i$ equals $n$. This ensures that both polynomials have degree $n$, since the first quotient is equal to $1$ by Remark~\ref{rem:first-last}(i).
	\item The sequence of constant terms is such that the constant terms of the two last remainders are respectively $1$ and $0$ (due to Remark~\ref{rem:first-last}(ii)), while the first two remainders (i.e., the reconstructed pair) must have constant term $1$.
\end{compactitem}

The intermediate terms, on the contrary, do not have any constraints and can be chosen freely. Thus, given the degree $n$ and the length $k$ of the quotients' sequence, enumerating the sequences of intermediate terms amount to enumerating all binary strings of length $n-k$, which are in total $I_{n,k} = 2^{n-k}$.

\section{Constant Terms Sequences as a Regular Language}
\label{sec:const}
The generic step $i$ of Euclid's algorithm applied to $(f,g) \in S_n$ corresponds to the Euclidean division:
\begin{equation}
	\label{eq:euc-div}
	r_i(X) = q_{i+1}(X)r_{i+1}(X) + r_{i+2}(X) \enspace ,
\end{equation}
where $r_i(x)$ and $r_{i+1}(X)$ are respectively the dividend and the divisor polynomial, $q_{i+1}(X)$ is the quotient, and $r_{i+2}(X)$ is the remainder of the division between $r_i(X)$ and $r_{i+1}(X)$. For $i=1$, one has $r_1(X) = f(X)$ and $r_2(X) = g(X)$. Then, the process is repeated by shifting the divisor to become the dividend, whereas the remainder becomes the divisor.

We interpret the presence or the absence of the constant terms in $r_i, r_{i+1}$ as the \emph{state} of a discrete dynamical system, described by a pair $(c_i, c_{i+1})$ where $c_i, c_{i+1} \in \F_2$ respectively denote the constant terms of $r_i$ and $r_{i+1}$. Remark that if $(f,g) \in S_n$ then $(c_i,c_{i+1}) \in (\F_2^2)^* = \{(1,1), (1,0), (0,1)\}$, since by Equation~\eqref{eq:euc-div} $(c_i,c_{i+1}) = (0,0)$ for any $i$ implies that $f$ and $g$ both have a null constant term.

\begin{figure}[t]
	\centering
	\begin{subfigure}{.5\textwidth}
		\centering
		\begin{tabular}{cc|c}
			\hline
			$(c_i,c_{i+1})$ & $s_{i+1}$ & $\delta((c_i,c_{i+1}),s_{i+1})$ \\
			\hline
			$(1,1)$        & $0$       & $(1,1)$                     \\
			$(1,1)$        & $1$       & $(1,0)$                     \\
			$(1,0)$        & $0$       & $(0,1)$                     \\
			$(1,0)$        & $1$       & $(0,1)$                     \\
			$(0,1)$        & $0$       & $(1,0)$                     \\
			$(0,1)$        & $1$       & $(1,1)$                     \\
			\hline 
		\end{tabular}
	\end{subfigure}%
	\begin{subfigure}{.5\textwidth}
		\centering
		\resizebox{!}{4cm}{
			\begin{tikzpicture}
				[->,auto,node distance=1.5cm, every loop/.style={min distance=12mm},
				empt node/.style={font=\sffamily,inner sep=0pt,outer sep=0pt},
				circ node/.style={circle,thick,draw,font=\sffamily\bfseries,minimum
					width=0.8cm, inner sep=0pt, outer sep=0pt}]
				
				\node [circ node] (n11) {$11$};
				\node [empt node] (e1) [below = 2.25cm of n11] {};
				\node [circ node] (n10) [right = 1.5cm of e1] {$10$};
				\node [circ node] (n01) [left = 1.5cm of e1] {$01$};
				
				\draw [->, thick, shorten >=0pt,shorten <=0pt,>=stealth] (n11) 
				edge[bend left=20] node (f5) [above right]{$1$} (n10);
				\draw[->, thick, shorten >=0pt,shorten <=0pt,>=stealth] (n11) edge[loop
				above] node (f3) [above]{$0$} ();
				\draw [->, thick, shorten >=0pt,shorten <=0pt,>=stealth] (n10) 
				edge[bend left=20] node (f5) [below]{$0/1$} (n01);
				\draw [->, thick, shorten >=0pt,shorten <=0pt,>=stealth] (n01) 
				edge[bend left=20] node (f5) [above]{$0$} (n10);
				\draw [->, thick, shorten >=0pt,shorten <=0pt,>=stealth] (n01) 
				edge[bend left=20] node (f5) [above left]{$1$} (n11);           
			\end{tikzpicture}
		}
	\end{subfigure}
	\caption{Transition table and graph realizing $\delta$.}
	\label{fig:delta}
\end{figure}
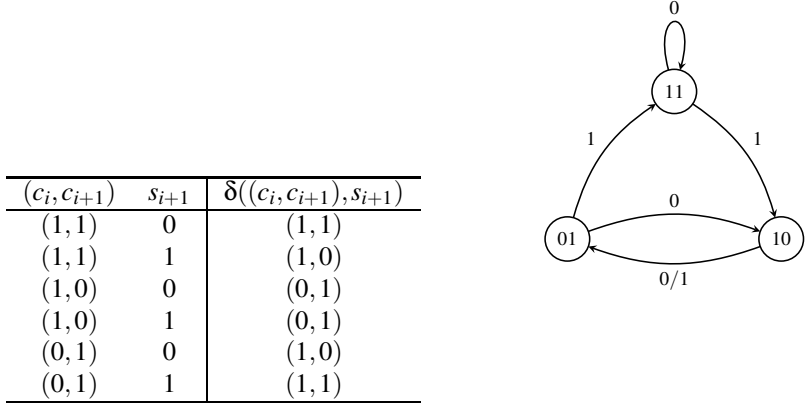

The transition function $\delta: (\F_2^2)^* \times \F_2 \rightarrow (\F_2^2)^*$ maps the pair $(c_i,c_{i+1})$ and the constant term $s_{i+1}$ of the $(i+1)$-th quotient to the next pair $(c_{i+1}, c_{i+2})$ by Equation~\eqref{eq:euc-div}. Figure~\ref{fig:delta} depicts the truth table and the transition graph of $\delta$.

We now consider this dynamical system as a \emph{Finite State Automaton} (FSA). The reason is that we can characterize the ``correct'' sequences of constant terms $s_i$ (i.e., those that give a pair $(f',g') \in A_n$ at the end of dilcuE's algorithm) as the words of the language recognized by the FSA. Hence, we need to define the initial and accepting states of the automaton. Given $(f,g) \in S_n^2$, the sequence $q_1,q_2,\cdots$ of quotients computed through Euclid's algorithm induces a path on the FSA graph. This path starts from the state $(1,1)$, and it is labelled by the constant terms $s_1,s_2,...$ of the quotients.

Notice that the FSA is \emph{permutative}: by taking two distinct states and reading the same constant term $s_{i+1}$, the two output states are distinct as well. This can be easily checked from the truth table of $\delta$. A simple induction argument shows that this property holds also for sequences of constant terms. Hence, if one starts from two different initial states and apply the same sequence of constant terms, the final states are also different. The inverse automaton, which corresponds to dilcuE's algorithm, is thus obtained by simply inverting the arrows in the transition graph of the original FSA. The initial state of this inverse automaton will be $(1,0)$, on account of Remark~\ref{rem:first-last}. For the accepting state, we should intuitively select $(1,1)$, since we want a pair of $A_n$ at the end of dilcuE's algorithm. Notice however that the first quotient in Euclid's algorithm (therefore, the last one in dilcuE's) is always $1$, due to Remark~\ref{rem:first-last}(i). Hence, we can shorten the sequence of quotients by one element, and append $1$ to it. Consequently, since the only way to reach $(1,1)$ in the inverse FSA by reading a $1$ is from $(1,0)$, we can also define $(1,0)$ as the only accepting state.

The classic state elimination method~\cite{hopcroft09} gives us the following regular expression for the language recognized by the inverse FSA:
\begin{equation}
	\label{eq:rl}
	L_r = (0(0+1)+(10^*1(0+1)))^* \enspace .
\end{equation}

We have thus obtained the following result:
\begin{lemma}
	\label{lm:reg-lang}
	The sequences of constant terms for the quotients visited by dilcuE's algorithm when generating a coprime pair $(f,g) \in A_n$ form a regular language $L_r$, whose regular expression is defined by Equation~\eqref{eq:rl}.
\end{lemma}

Hence, enumerating the sequences $s_1,\cdots,s_k$ is equivalent to generating all words of length $k$ in $L_r$. Several algorithms are available for this task, see e.g.~\cite{makinen97}. To count the number of words, we apply the \emph{Chomsky-Sch\"{u}tzenberger enumeration theorem}~\cite{chomsky59}, which uniquely identifies a regular language $L$ with a rational Formal Power Series (FPS) $\mathcal{F}_L = \sum_{k=0}^{\infty} \ell_kX^k$. This leads us to the following:

\begin{lemma}
	\label{lm:cf}
	The generating function $G(X)$ for the FPS associated to $L_r$ and the recurrence for the number $\ell_k$ of words of length $k \in \N$ in $L_r$ are given by:
	
	\begin{equation}
		\label{eq:cf}
		G(X) =  \frac{1-X}{1-X-2X^2} \enspace , \enspace \ell_k = \frac{2^k + 2\cdot(-1)^k}{3} \enspace .
	\end{equation}
	\begin{proof}
		Omitted (see~\cite{formenti22}).
	\end{proof}
\end{lemma}

\section{Sequences of Quotients Degrees and Compositions}
\label{sec:comp}
The only constraint enforced on the sequences of quotients' degrees $d_1,\cdots, d_k$ is that they must sum to the final degree $n$. The order of the summands is relevant: permuting the degrees gives rise to a different sequence of quotients. Hence, we want to enumerate and count the number of ways in which $n$ is obtained as an ordered sum of $k$ natural numbers. These are also known as $k$-\emph{compositions} of $n \in \N$ in combinatorics~\cite{riordan12}, and a simple way to represent them is by means of $n-1$ \emph{boxes} interleaved by $n$ occurrences of $1$:
\[
1 \overbrace{\square 1 \square \ldots \square 1 \square}^{n-1} 1 \enspace ,
\]
where each box can be either a comma (,) or a plus (+). A comma separates two different parts in a composition, while a plus adds two adjacent 1s together. Notice that we cannot take the composition where all boxes are set to $+$, since a sequence composed of just one quotient cannot occur in dilcuE's algorithm.

Once the length $k$ of the sequence of quotients is fixed, generating the corresponding degrees is equivalent to the enumeration of all binary strings of length $n-1$ with $k-1$ ones in them, which can be accomplished, for instance, by one of the several  algorithms described by Knuth~\cite{knuth11}. Further, the number of all compositions of $n$ of length $k$ is given by the binomial coefficient $\binom{n-1}{k-1}$. Therefore, we obtained the following result:
\begin{lemma}
	\label{lm:comp}
	The number of sequences of degrees $d_1,\cdots,d_k$ of the final degree $n \in \N$ for the quotients visited by dilcuE's algorithm is:
	\begin{equation}
		\label{eq:comp}
		D_{n,k} = \binom{n-1}{k-1} \enspace .
	\end{equation}
\end{lemma}

Putting together the results presented above, the following pseudocode exhaustively enumerates all pairs of coprime polynomials in $A_n$ for a given input degree $n$ and quotients' sequence length $2 \le k \le n$:
\begin{compactitem}
	\item For each composition $comp$ of $n$ of length $k$ do:
	\begin{compactitem}
		\item[(1)] Generate the degrees' sequence $deg$ corresponding to $comp$
		\item[(2)] For each intermediate terms sequence $seq$ do:
		\begin{compactitem}
			\item[(2a)] Adjoin $seq$ to $deg$ to get a quotients's sequence $quot$
			\item[(2b)] For each constant term sequence $const$ of length $k$ do:
			\begin{compactitem}
				\item Adjoin $const$ to $quot$
				\item Apply DilcuE's algorithm from $(1,0)$ by using the sequence $quot$
			\end{compactitem}
		\end{compactitem}
	\end{compactitem}
\end{compactitem}

Finally, the Lemma below provides a different derivation for the number $a_n = |A_n|$ of coprime polynomials of degree $n$ and with a nonzero constant term.
\begin{lemma}
	\label{lm:count}
	The number of pairs of coprime polynomials of degree $n$ with nonzero constant term is equal to:
	\begin{equation}
		\label{eq:count}
		a_n = \sum_{k=2}^n 2^{n-k} \cdot \binom{n-1}{k-1} \cdot \frac{2^k + 2\cdot(-1)^k}{3} = 2 \cdot \frac{4^{n-1}-1}{3} \enspace .
	\end{equation}
	\begin{proof}
		Omitted (see~\cite{formenti22}).
	\end{proof}
\end{lemma}

\bibliographystyle{splncs04}
\bibliography{bibliography_preprint}

\end{document}